%% file: charm.tex
\documentclass[conference,10pt,letterpaper,nofonttune]{IEEEtran}
\usepackage[noend]{algpseudocode}
\usepackage[shortlabels]{enumitem}
\usepackage{acronym}
\usepackage{algorithm}
\usepackage{amsmath}
\usepackage[capitalize]{cleveref}
\usepackage{comment}
\usepackage{graphicx}
\usepackage{url}
\usepackage{tikz}
\usepackage{tikz-layers}
\usepackage{xspace}
\usepackage{flushend}

\usetikzlibrary{fit,arrows,calc}

\acrodef{ABS}{Almost Blank Subframe}
\acrodef{BS}{Base Station}
\acrodef{BSU}{Base Station Unit}
\acrodef{CBRS}{Citizen Broadband Radio Service}
\acrodef{CNN}{Convolutional Neural Network}
\acrodef{COTS}{Commercial Off-The-Shelf}
\acrodef{ChARM}{Channel Aware Reactive Mechanism}
\acrodef{ChASM}{Channel Adaptive Selection Mechanism}
\acrodef{DL}{Deep Learning}
\acrodef{DNN}{Deep Neural Network}
\acrodef{EARFCN}{EUTRA Absolute Radio-Frequency Channel Number}
\acrodef{EPC}{Evolved Packet Core}
\acrodef{ES}{Entropy Selection}
\acrodef{FPGA}{Field Programmable Gate Array}
\acrodef{ICIC}{Inter-Cell Interference Coordination}
\acrodef{ISM}{Industrial Scientific Medical}
\acrodef{LBT}{Listen Before Talk}
\acrodef{MCHEM}{Massive Channel Emulator}
\acrodef{PRU}{Policy Reacting Unit}
\acrodef{QoS}{Quality of Service}
\acrodef{RN}{Residual Network}
\acrodef{SDR}{Software Defined Radio}
\acrodef{SNR}{Signal-to-Noise-Ratio}
\acrodef{SCU}{Spectrum Classification Unit}
\acrodef{SoA}{State of Art}
\acrodef{SoC}{System on Chip}
\acrodef{SpectroCell}{Spectropathic Cell}
\acrodef{UE}{User Equipment}
\acrodef{FCC}{Federal Communication Commission}
\acrodef{MNO}{Mobile Network Operator}
\acrodef{PDU}{Policy Decision Unit}
\acrodef{RIC}{RAN Intelligent Controller}
\acrodef{RU}{Radio Unit}
\acrodef{DU}{Distributed Unit}
\acrodef{CU}{Central Unit}
\acrodef{PHY}{Physical layer}

\newcommand{\stdimg}[3]{\begin{figure} \centering \includegraphics[width=\columnwidth]{img/#1}  \caption{#3} \label{#2} \end{figure}}

\newcommand{\FW}{\texttt{ChARM}\xspace}

\definecolor{pink}{RGB}{215,25,28}
\definecolor{lavander}{RGB}{253,174,97}
\definecolor{yellow}{RGB}{255,255,191}
\definecolor{green}{RGB}{171,221,164}
\definecolor{blue}{RGB}{43,131,186}

\title{\FW: NextG Spectrum Sharing Through Data-Driven Real-Time O-RAN Dynamic Control} %\vspace{-1.2cm}}

\author{\IEEEauthorblockN{Luca Baldesi, Francesco Restuccia, and Tommaso Melodia }
\IEEEauthorblockA{Institute for the Wireless Internet of Things, Northeastern University, United States}
\IEEEauthorblockA{Email: \{l.baldesi,\ frestuc, \ melodia\}@northeastern.edu\vspace{-1cm}}
\thanks{This article is based on material supported in part by the US National Science Foundations under the grants CNS \#1923789 and \#1925601, and by the US Office of Naval Research under grant ONR N00014-20-1-2132.}}

\IEEEoverridecommandlockouts  % to allow the thanks command

\begin{document}

\maketitle

\begin{abstract}
Today's radio access networks (RANs) are monolithic entities which often operate statically on a given set of parameters for the entirety of their operations.
To implement realistic and effective spectrum sharing policies, RANs will need to seamlessly and intelligently change their operational parameters. In stark contrast with existing paradigms, the new O-RAN  architectures for 5G-and-beyond networks (NextG) separate the logic that controls the RAN from its hardware substrate, allowing unprecedented real-time fine-grained control of RAN components. In this context, we propose the Channel-Aware Reactive Mechanism (\FW), a data-driven O-RAN-compliant framework that allows (i) sensing the spectrum to infer the presence of interference and (ii) reacting in real time by switching the distributed unit (DU) and radio unit (RU) operational parameters according to a specified spectrum access policy. \FW is based on neural networks operating directly on unprocessed I/Q waveforms to determine the current spectrum context. \FW does not require any modification to the existing 3GPP standards. It is designed to operate within the O-RAN specifications, and can be used in conjunction with other spectrum sharing mechanisms (e.g., LTE-U, LTE-LAA or MulteFire). We demonstrate the performance of \FW in the context of spectrum sharing among LTE and Wi-Fi in unlicensed bands, where a controller operating over a RAN Intelligent Controller (RIC) senses the spectrum and switches cell frequency to avoid Wi-Fi.
We develop a prototype of \FW using srsRAN, and leverage the Colosseum channel emulator to collect a large-scale waveform dataset to train our neural networks with. To collect standard-compliant Wi-Fi data, we extended the Colosseum testbed using system-on-chip (SoC) boards  running a modified version of the OpenWiFi architecture.
Experimental results show that \FW achieves accuracy of up to 96\% on Colosseum and 85\% on an over-the-air testbed, demonstrating the capacity of \FW to exploit the considered spectrum channels.
\end{abstract}

\section{Introduction and Motivation}
\label{sec:intro}

According to the new Cisco Annual Internet Report, 5G and beyond (NextG) networks will support more than 10\% of the world's mobile connections by 2023, with more than 5.7B users  -- 70\% of the global population -- using mobile cellular connectivity \cite{cisco}. Due to this sheer growth in wireless demand, current spectrum bands below 6~$\mathrm{GHz}$ will inevitably become saturated. For this reason, the \ac{FCC} has recently opened 1.2~$\mathrm{GHz}$ of spectrum in the 6~$\mathrm{GHz}$ band, basically quadrupling the amount of space available for routers and other unlicensed devices \cite{FCC6GHz}. Moreover, 150 MHz of spectrum in the \ac{CBRS} band can now be accessed~\cite{tarver_enabling_2019,federal2015amendment}, shared with incumbent radar communications. 

As new spectrum bands become open for unlicensed usage, it becomes crucial to protect incumbent users (\textit{i.e.}, previous license owners), as well as establishing fair coexistence among unlicensed users.  For example, it has been demonstrated that Wi-Fi throughput can drop up to 70\% without a dedicated LTE co-existence mechanism~\cite{bocanegra_e-fi_2019}. To this end,  \textit{spectrum sharing} has emerged as a key technology to fuel wireless growth in these bands \cite{WiFi5GCoexist}. Spectrum sharing enables multiple categories of users to opportunistically select frequencies and bandwidth of operation, according to given constraints (e.g., band limits and incumbent priorities). 

Due to the dynamic nature of spectrum policies and the unpredictability of unlicensed usage, spectrum sharing will require radio access networks (RANs) to change their operational parameters intelligently and according to the current spectrum context. Although existing RANs do not allow  real-time reconfiguration, the fast-paced rise of the Open RAN movement and of the O-RAN framework \cite{O-RAN} for 5G-and-beyond (NextG) networks, where the hardware and software portions of the RAN are logically disaggregated, will allow seamless reconfiguration and optimization of the radio components \cite{bonati2020open}. Despite their compelling necessity, to the best of our knowledge there are no O-RAN-ready technologies that can drive real-time RAN optimization, as discussed in details in Section \ref{sec:soa}.

\begin{figure}[!h]
\centering
\includegraphics[width=\columnwidth]{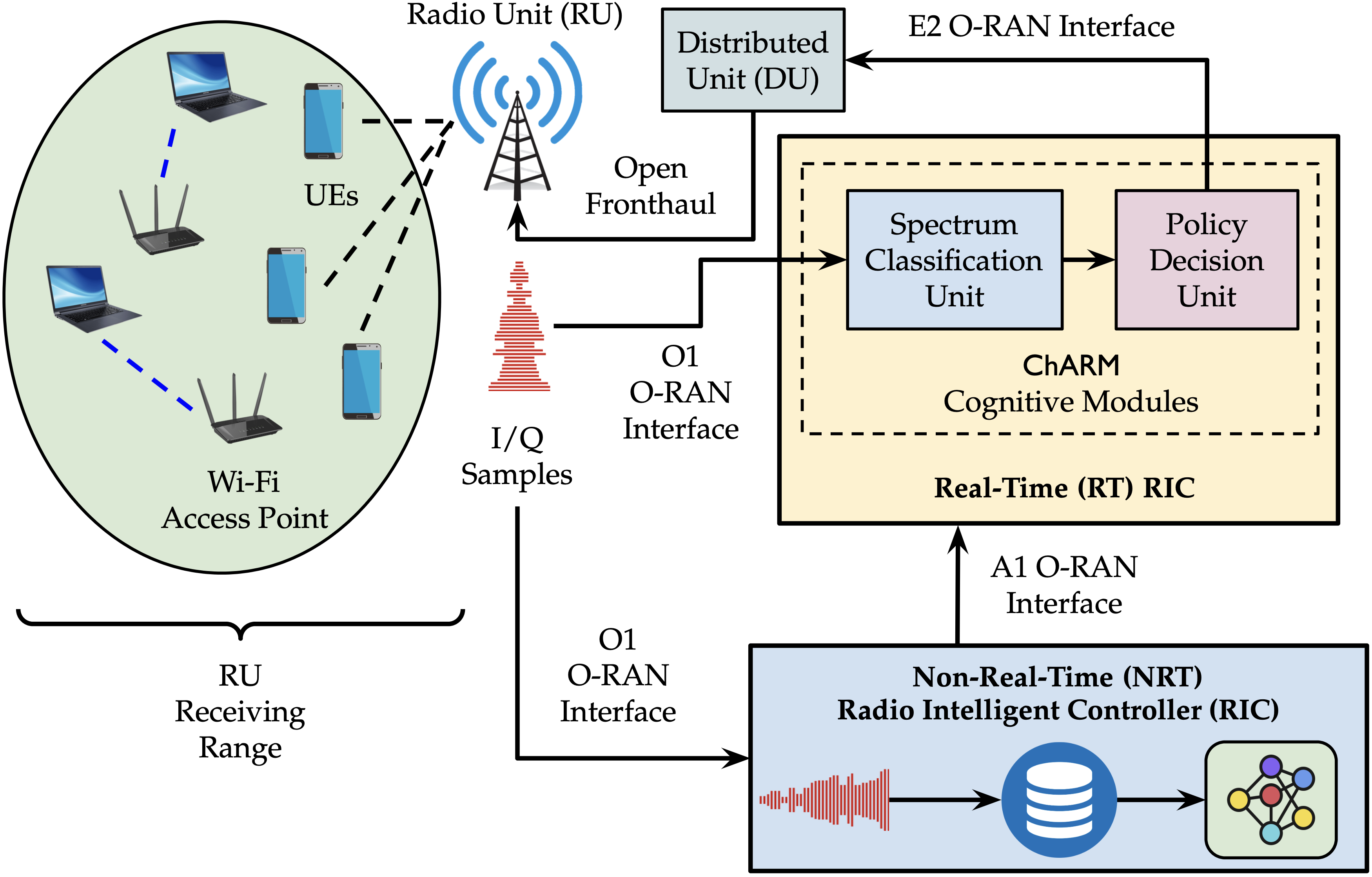}
\caption{Overview of the O-RAN-based \FW spectrum-sharing framework.\vspace{-0.5cm}}
\label{fig:charm_overview}
\end{figure}

For this reason, in this paper we propose the \textit{Channel-Aware Reacting Mechanism} framework (in short, \FW).  Fig.~\ref{fig:charm_overview} shows a high-level overview of \FW and its main logical components, including the O-RAN interfaces used to collect and exchange data among the different components. \FW is a data-driven framework that enables RAN owners to (i) sense the spectrum to understand the current context through a \ac{SCU}; (ii) react in real time by switching the \ac{DU} and \ac{RU} operational parameters according to a specified spectrum access policy decided by \ac{PDU}. Both \ac{SCU} and \ac{PDU} are located in the O-RAN near-real-time \ac{RIC}, which receives input by the non-real-time \ac{RIC}. The latter is tasked with (i) collecting the spectrum I/Q data and creating a dataset; (ii) training and testing the machine learning (ML) algorithms that are eventually deployed onto the real-time \ac{RIC} through the A1 interface. 

\begin{figure}%[!h]
\centering
\includegraphics[width=\columnwidth]{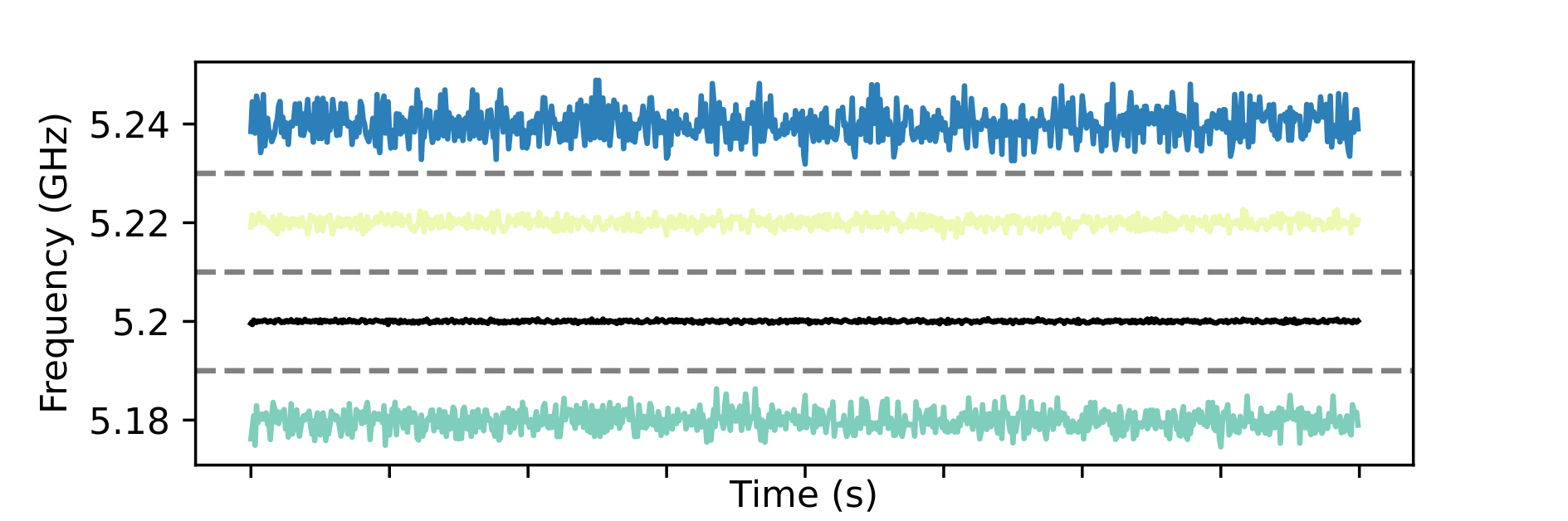}
\caption{Spectrum sensing based on different frequencies. The sensing bandwidth is set to 20 MHz (the LTE channel bandwidth in the \ac{ISM} band). Some channels present radio signals, but their identification as legitimate communication or just noise requires a classifier.}
\label{fig:channels}
\end{figure}

The key innovation behind \FW is providing Open RAN networks with the capability to intelligently determine which wireless technology is utilizing the spectrum, so that intelligent spectrum policies can be implemented. To this end, the  \ac{SCU} of \FW leverages \ac{DNN} trained on unprocessed I/Q samples to classify communication technologies with low latency~\cite{yang_blind_2020}. Different from prior work, however, we design our classifiers to include an \textit{abstain class} (see, for example, \cite{liu_deep_2019}) to minimize misclassifications of unknown wireless technologies (something likely in the \ac{ISM} band). \Cref{fig:channels} shows an example of spectrum occupation in the \ac{ISM} band between 5.18 and 5.24 GHz, where different wireless technologies are utilizing the spectrum. According to the given spectrum  utilization rule, the \ac{PDU} unit of \FW may decide to switch to the empty 5.2 GHz band, also called inter-channel sharing, or activate a co-existence mechanism inside the occupied channel, such as (LTE-U, LTE-LAA or MulteFire). This methodology is called intra-channel sharing. 

To the best of our knowledge, ours is the first framework providing the capabilities defined above to O-RAN-ready networks. The closest work to ours is due to Tarver \textit{et al.}~\cite{tarver_enabling_2019}, who presented a solution for sensing and reacting nodes for the \ac{CBRS} context, as well as Uyadov \textit{et al.}~\cite{uyadov_deepsense_nodate}, who propose a sensing and reacting framework optimizing the usage of fragmented, unused portions (holes) of spectrum. However, these approaches require deep modifications of the 3GPP and 802.11 standards, which ultimately makes them not readily adaptable to state-of-the-art O-RAN networks. Some solutions \cite{qian_multi-operator_2021,mosleh_dynamic_2020} rely on a centralized orchestrating node, do not actively sense the state of the spectrum, or are inherently limited to two technologies (LTE and WiFi). Moreover, legacy approaches do not allow the customization of the behavior by the \ac{MNO}, and are not compatible with O-RAN specifications. Conversely, operators should be able to specify customized reactions tailored to the sensed technology and the band of operation.

As part of the novel contributions of this paper, we address (i) the need for a large waveform dataset to train the \ac{DNN} with, and (ii) the development of a real-time working prototype. To experiment in both emulated and over-the-air channels, we develop a prototype for both the Colosseum channel emulator and the over-the-air Arena testbed. Colosseum enables researchers and practitioners to control the wireless channel environment while using state-of-the-art \ac{SDR} devices. While Colosseum has not been designed to work with Wi-Fi devices, we extend Colosseum with new hardware in the loop, proving its extreme flexibility and extensibility.
Our prototypes prove that  \FW is fully O-RAN-ready, it can interact with the 3GPP and 802.11 standards, and it is designed to be used in combination with any other intra-channel mechanisms. 

To summarize, this paper makes the following novel technical contributions:\smallskip

$\bullet$ We present \FW, an O-RAN-based framework for spectrum sharing in the \ac{ISM} band. \FW is composed by (i) a spectrum classification unit (SCU) based on \ac{DNN}s for real-time spectrum classification, (ii) a policy decision unit (PDU) that defines the actions to be taken upon the inference produced by the SCU;  \smallskip

$\bullet$ We design and implement a \FW prototype   based on standard-compliant srsRAN software. Through this prototype, we demonstrate \FW in the context of spectrum sharing among LTE and Wi-Fi in unlicensed bands, where the RU reactively switches cell frequency to avoid Wi-Fi according to the DNN-based SCU inference. We leverage the Colosseum channel emulator to collect a large-scale waveform dataset to train our neural networks with. To collect standard-compliant Wi-Fi data, we extended the Colosseum testbed using \ac{SoC} boards, running our patched version of OpenWiFi \cite{jiao2020openwifi}, an 802.11a/g/n implementation specifically designed for \ac{SoC} boards. We demonstrate the feasibility of our approach by deploying our software and the \ac{DNN} model, trained on Colosseum, in a wireless test-bed, Arena~\cite{bertizzolo_arena_2020}, and operating it in the \ac{ISM} band with incumbent WiFi communications. Experimental results show that our neural networks achieve accuracy of up to 96\% on Colosseum and 85\% on Arena, demonstrating the capacity of \FW to exploit the considered spectrum channels; \smallskip

$\bullet$ For reproducibility purposes and to stimulate further research, we provide access to our code and dataset (\cref{sec:conclusions}).

\section{Related Work}
\label{sec:soa}

A significant amount of prior work has tackled spectrum sharing in the \ac{ISM} band, primarily targeting spectrum sharing between LTE and WiFi.
Some approaches assume a collaboration between LTE and WiFi nodes; Chen et al.~\cite{chen_optimizing_2016} envision the creation of a LTE/WiFi super node, internally optimizing the spectrum usage fairness. Gawlowicz et al.~\cite{gawlowicz_enabling_2018} design a framework for side channel communication between WiFi access points and LTE \ac{BS}s.
These approaches, along with the one by Bocanegra et al.~\cite{bocanegra_e-fi_2019} that modifies the WiFi access point software, are challenging to deploy in practice, and hardly extensible to consider other technologies beyond LTE and WiFi. Some prior approaches achieve co-existence at the physical layer (PHY).  The work by Yun et al.~\cite{yun_supporting_2015} focuses on interference cancellation and beamforming exploiting multiple radio antennas. Almeida et al.~\cite{almeida_enabling_2013} focus on exploiting a 3GPP standard feature, the \ac{ABS}, and they paved the way for the standardization of LTE-U~\cite{qualcomm_technologies_inc_qualcomm_nodate}, originally proposed by Qualcomm, as a mean of intra-channel LTE co-existence. Guan and Melodia~\cite{guan_cu-lte_2016} mathematically modeled the fairness of LTE-U systems and proposed algorithms to maximize throughput under fairness constraints.

The solutions based on LTE-U could not be deployed in Europe and Japan, where regulations impose to use a \ac{LBT} mechanism (CSMA/CA-like) to access the \ac{ISM} band.
Hence, 3GPP standardized another technique called LTE-LAA~\cite{3gpp_r13_study_nodate}, which is an extension to LTE enabling \ac{LBT}.
Several works stemmed from this standardization effort; Challita et al.~\cite{challita_proactive_2018}, and later, Tan et al.~\cite{tan_qos-aware_2019}, propose to employ ML to forecast Wi-Fi transmission and optimize LTE consequently. Garcia Saavedra et al.~\cite{garcia-saavedra_orlaolaa_2018} raised attention on LTE-LAA unfairness cases and propose optimizing parameters to minimize them; Gao and Roy~\cite{gao_achieving_2020} addressed instead the unfairness by modeling LTE-LAA communications with Markov models.
The works by Chai et al.~\cite{chai_lte_2016} and Saha et al.~\cite{saha_demilte_2019} introduce modifications to the LTE base station and the WiFi access point, respectively.
Both these solutions and the ones based on LTE-U and LTE-LAA focus on intra-channel spectrum sharing.
Huang et al.~\cite{huang_deep_2020}, instead, propose to achieve a fair co-existence between LTE and WiFi transmission by inter-channel optimization through a real-time intensive CUDA computation.
Qian et al.~\cite{qian_multi-operator_2021} address the problem of centralized spectrum allocation among different \acp{MNO}. The approach by Mosleh et al.~\cite{mosleh_dynamic_2020} presents an ML framework to optimize the spectrum usage by LTE and WiFi. However, it does not include sensing functions and its application is limited to those two technologies. Even though existing work tackles wireless technology classification through \ac{DNN}~\cite{yang_blind_2020}, to the best of our knowledge, we are the first to propose a full-fledged O-RAN based framework for sensing and reacting cells, while maintaining full compatibility with the 3GPP standard.

\section{The \FW framework}\label{sec:framework}

\subsection{Background on O-RAN}

O-RAN and the NextG architecture are based on the 3GPP functional split. The functionalities of the base stations are virtualized and disaggregated, often running on multiple physical nodes. These functionalities are grouped in \ac{CU}, \ac{DU}, and \ac{RU}. Specifically, while \ac{CU} deals with protocols higher in the stack, \ac{DU} is responsible for time-critical operations (including most baseband processing), while the \ac{RU} is in charge of  radio frequency and of some \ac{PHY} functionalities (e.g., beamforming, fast Fourier transforms).

Moreover, O-RAN has been designed to embrace programmatic control based on ML and on the open source paradigm. For this reason, it exposes analytics and control knobs through the non real time \ac{RIC} and the near real time \ac{RIC}. These two components are responsible of the intelligent control of the network. The former handles operations with coarse time granularity (such as training a \ac{DNN} model, orchestration of containers, among others), while the latter handles operations that need to be performed within a second, for example, the inference of a \ac{DNN} model. The near real time \ac{RIC} also allows running customized network functions (called xApps), which \ac{MNO} can install in their nodes. \FW has been specifically tailored to be deployed as an xApp in the near real time \ac{RIC} and integrated in the NextG architecture.\vspace{-0.1cm}

\subsection{Overview of \FW}

\begin{figure}[!h]
	\centering
	\includegraphics[width=\columnwidth]{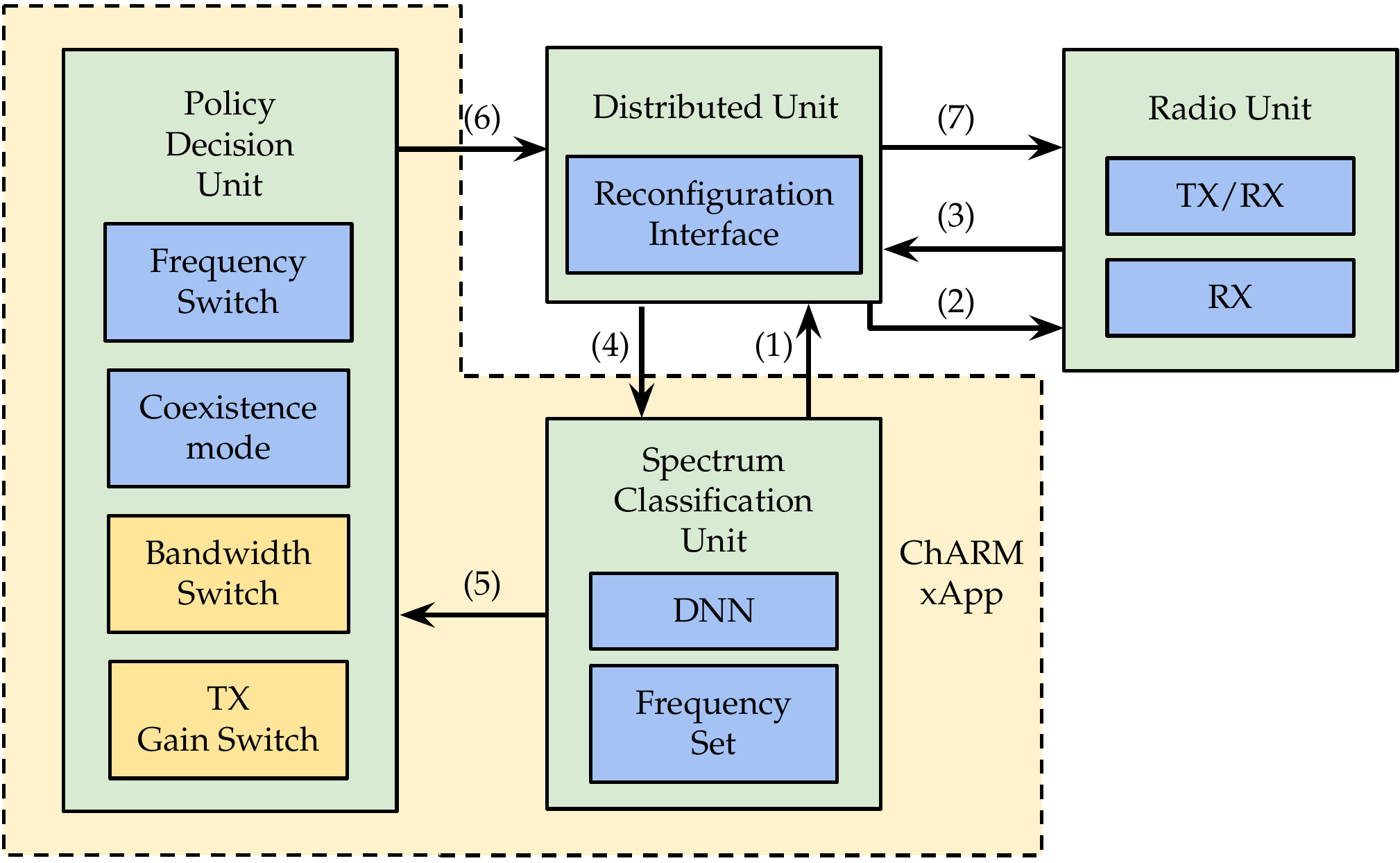}
    % original figure: https://docs.google.com/drawings/d/1_rCYtDcxlP9cbt1lRTE-1rVnyJGMxYpU3SAGpC0NhKs/edit
	\caption{\FW system framework. Sub-blocks in the Policy Decision Unit indicate possible reacting strategies. While the framework is flexible enough to implement several different ones, for the purpose of this paper, we employ those highlighted in blue.}
	\label{fig:sts}
\end{figure}

\Cref{fig:sts} represents a high-level overview of the main logical components of \FW in the context of O-RAN. The framework requires at least two co-located radios, one  for mobile network communications (indicated with TX/RX),  and another for sensing (indicated with RX). Moreover, \FW is composed of (i) a spectrum classification unit (\ac{SCU}) responsible of scanning various given frequencies and classify each of them, which comprises of a pre-defined set of frequencies to evaluate and a \ac{DNN} for I/Q sample classification, (ii) a policy decision unit (\ac{PDU}) which takes as input the latest frequency evaluation by the classifier, embeds a policy which can be customized by the operator (see \cref{sec:policy}), and communicates to the \ac{DU} unit the changes to apply to the on-going communication, and (iii) the \ac{DU}, which implements the control interface to receive commands from the \ac{PDU}. 

\textbf{A walk-through.}~We provide an overview of the key operations of \FW with the help of \cref{fig:sts}. While the \ac{RU} can be communicating with zero or more \acp{UE},  the \ac{SCU} periodically indicates to the \ac{DU} (step 1) to reconfigure the RX radio to a different frequency (step 2).
Then, the \ac{DU} collects I/Q samples (step 3), which are fed to the \ac{SCU} (step 4). Then, the \ac{SCU} classifies the samples through the \ac{DNN} and the result (i.e., frequency and class) is provided to the \ac{PDU} (step 5). The \ac{PDU} is thus aware of (i) which frequency the \ac{RU} is using for mobile communication, (ii) which is its latest assigned class, and (iii) the classes assigned to the other frequencies under sensing. 
The \ac{PDU} may react to the sensed spectrum state triggering one or multiple of its functionalities, for example:

\begin{itemize}
    \item Frequency switch, which makes the \ac{RU} change center frequency;
    \item Coexistence mode, which enables or disables a specific coexistence mechanism of the \ac{RU};
    \item Bandwidth switch, which changes the signal bandwidth in the \ac{ISM} band;
    \item TX gain switch, changes transmission gain of the \ac{RU}.
\end{itemize}

The chosen reacting functions depend on the sensed spectrum state and the network operator policy (detailed in \cref{sec:policy}), and they are sent to the \ac{DU} (step 6). The \ac{DU} adapts the spectrum usage with respect to the received commands (step 7). In the case of frequency or bandwidth switch, it communicates with the \ac{UE}s through a 3GPP-standard compliant reconfiguration message~\cite{3gpp_r10_radio_nodate} to grant the continuity of the ongoing communications.\vspace{-0.1cm}

\subsection{\acl{SCU}}
\label{sec:scu}

\textbf{Sensing Procedures.}
Sampling a given frequency implies tuning the receiving radio and wait for the phase locked loop (PLL) to stabilize. This can take up to several tenths of seconds for each single channel to inspect. Alternatively, \ac{SDR}s can be used to sense a larger portion of spectrum (multiple of channel width) and then filter out the channels of interest. While the latter does not present the inconvenience of frequency retuning, it has two main drawbacks: (i)  state-of-the-art filtering, the polyphase channelizer~\cite{polyphase}, requires a large numbers of taps to be accurate, at the cost of being slower than retuning, and (ii) \ac{SDR} maximum input bandwidth is constrained by hardware (e.g., 80 MHz on Ettus Devices USRP X310), which limits the sensing capabilities. Early experiments -- not included due to space limitations -- have shown the impracticability of the channelizer solution. For these reasons, \FW employs a frequency hopping sensing mechanism.

\textbf{DNN.}~I/Q samples represent a time series stream of data. Existing work has proven that \ac{CNN}s are suitable for mining recurrent patterns and identifying key features in the wireless domain. \ac{CNN}s have been used extensively for modulation and spectrum classification~\cite{Oshea2016convolutional,uyadov_deepsense_nodate}.
However, in the computer vision field~\cite{resnet_CV}, and later in the audio processing~\cite{resnet_wave}, the concept of deep \ac{RN} has emerged and has been demonstrated to be increasingly effective. For example, \acp{RN} use convolutional layers and bypass connections, allowing the stacking of significant amounts of layers and the consequent effective analysis of data at many different scales.
For this reason \ac{RN}s have been applied to I/Q stream analysis too~\cite{resnet_modulation}, and are considered in this paper.

\subsection{\acl{PDU}}
\label{sec:policy}
The goal of the \ac{PDU} is to periodically collect the latest information generated by the classifier and, according to a given policy, instruct the \ac{DU} on which spectrum changes to undertake.
The policy is defined by an \ac{MNO} to customize the \ac{PDU} decisions, and it is bundled in the xApp.
It is implemented as a function evaluating the current system state, defined by the classes assigned to the frequencies under evaluation, and the current communication frequency.

\Cref{alg:pdu} presents the periodic routine run by the \acl{PDU}.
Specifically, \textit{ch\_classes} is an associative map, assigning to each sensed frequency by the \ac{SCU} a technology label (e.g., $\{5.18\rightarrow \text{Clear},5.20\rightarrow \text{LTE},5.22\rightarrow \text{Unknown}\}$) generated by the \ac{DNN}. The \ac{PDU} periodic routine calls the policy function to determine the actions to perform. If the policy dictates a change of parameters, it triggers the respective operations of the \ac{DU} reconfiguration interface.

\begin{algorithm}
\footnotesize
\caption{Periodically run \ac{PDU} algorithm}
	\label{alg:pdu}
\begin{algorithmic}[1]

\Procedure{PRI\_update}{$ch\_classes,curr\_freq$}
	\State $freq,coex,pw,bw \leftarrow policy(ch\_classes,curr\_freq)$
	\If{$curr\_freq \neq freq$}
	    \State handover($freq$)
	    	\EndIf
	    \State set\_coexistence($coex$)
	    \State set\_tx\_power($pw$)
	    \State set\_bw($bw$)

\EndProcedure

\end{algorithmic}
\end{algorithm}

\textbf{Frequency switch.}
\FW performs a handover whenever the \ac{PDU} decides to change frequency. In this phase, it is crucial to guarantee  continuity of the session and avoid disconnections of mobile \ac{UE}s.  3GPP standards already indicate the procedure for inter-frequency handovers, and \FW exploits it to grant standard compliant seamless handovers. \FW \ac{RU} manages two cells, one of them serving the \ac{UE}s, while the other is kept idle. When \FW changes operating frequency, (i) it changes the frequency of the idle cell with the target, and (ii) its \ac{DU} sends a message handover through a RRC Reconfiguration Message~\cite{3gpp_r10_radio_nodate} to the \ac{UE}s.

\textbf{Co-existence mode.}
\FW targets spectrum sharing optimization both at the inter-channel level and at the intra-channel level. \FW can hence work in two modes, co-existing and non co-existing. In non co-existing mode the \ac{PDU} makes the network nodes communicate the regular way.
When the \ac{PDU} dictates a handover to a frequency already occupied, it can switch the \ac{DU} to activate a predefined co-existence technique for the detected incumbent technology.
Possible mechanisms include: the increase of \acl{ABS} periods~\cite{almeida_enabling_2013,guan_cu-lte_2016,cano_fair_2017} for the co-existence with WiFi, and the establishment of an X2 interface and the subsequent coordination through 3GPP \ac{ICIC} techniques~\cite{ICIC_survey} for co-existence with LTE \acp{BS}.
However, the specific choices for intra-channel spectrum sharing algorithms and performance are out of the scope of this paper, and we simply assume that, whenever a co-existence mode is required and activated, a sensible co-existence mechanism choice is set in place and the performance improves.

\section{ \FW Prototype}
\label{sec:prototype}

We first describe in Section \ref{sec:use-case} the use-case scenario of \FW we consider, as well as its design and implementation.

\subsection{Use-case Scenario: Spectrum Sharing in ISM Bands}\label{sec:use-case}

\begin{figure}[!h]
    \centering
    \includegraphics[width=\columnwidth]{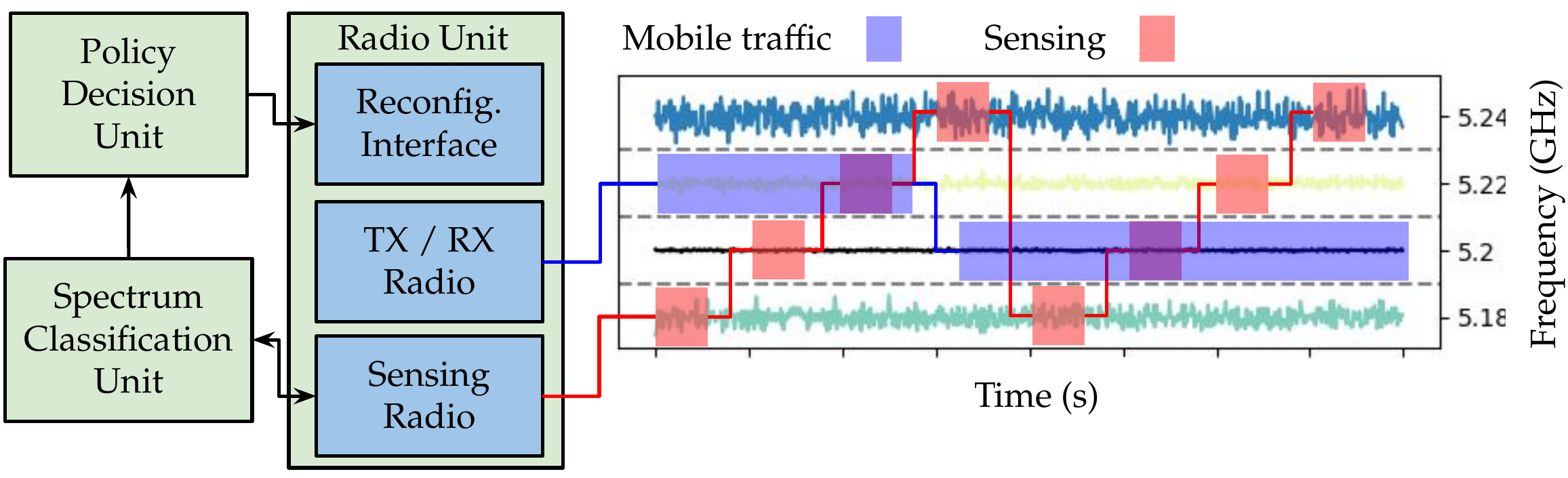}
    \caption{View of the \FW prototype components and their interactions.}
    \label{fig:charm_srsran}
    %Original: https://docs.google.com/drawings/d/1LsWKMuvsW7Yg5yqo8nH-yjMkFu_Zmve9Pq_S6hZk1bo/edit
\end{figure}

We cast \FW in the context of spectrum sharing in the license-free industrial, scientific and medical (ISM) bands, where a 5G O-RAN cellular network (hereafter referred to as LTE for simplicity), Wi-Fi users and incumbent spectrum licensees need to share the same spectrum and thus coexist with each other. \Cref{fig:charm_srsran} depicts the components of the \FW prototype and their main interactions. 
We depict with a shade of blue and red, respectively, the interactions of \FW with the channel: mobile communication and sensing. 
The image illustrates the inter-frequency spectrum optimization introduced in \cref{sec:intro}; \FW addresses that challenge by dynamically reconfiguring the mobile traffic to handover to the unoccupied sensed frequency.

In this scenario, the radio unit (RU) is composed by a reconfiguration interface, a sensing radio and a TX/RX radio. The sensing radio periodically listens to the channel and feeds the received waveform to the spectrum classification unit (SCU), which then sends its inference to the policy decision unit (PDU). The latter then interacts with the RU through the interface, which lets the TX/RX radio switch channel according to a given policy. In our experiments, we use a policy function based on a ranking of traffic classes.
The two rankings we use in the shown experiments are presented in in \cref{tab:policy}. 

\begin{table}[!h]
	\caption{Class rankings to be used for policy. \\Higher values imply higher preference.}
	\centering
	(a)
	\begin{tabular}{|c|c|}
	\hline
	Clear & 3\\
	\hline
	WiFi & 2\\
	\hline
	LTE & 1\\
	\hline
	Unknown & 0\\
	\hline
	\end{tabular}
	~~~~(b)
	\begin{tabular}{|c|c|}
	\hline
	Clear & 3\\
	\hline
	WiFi & 1\\
	\hline
	LTE & 2\\
	\hline
	Unknown & 0\\
	\hline
	\end{tabular}
	\label{tab:policy}
\end{table}

\begin{figure}[h!]
	\centering
	\includegraphics[width=0.9\columnwidth]{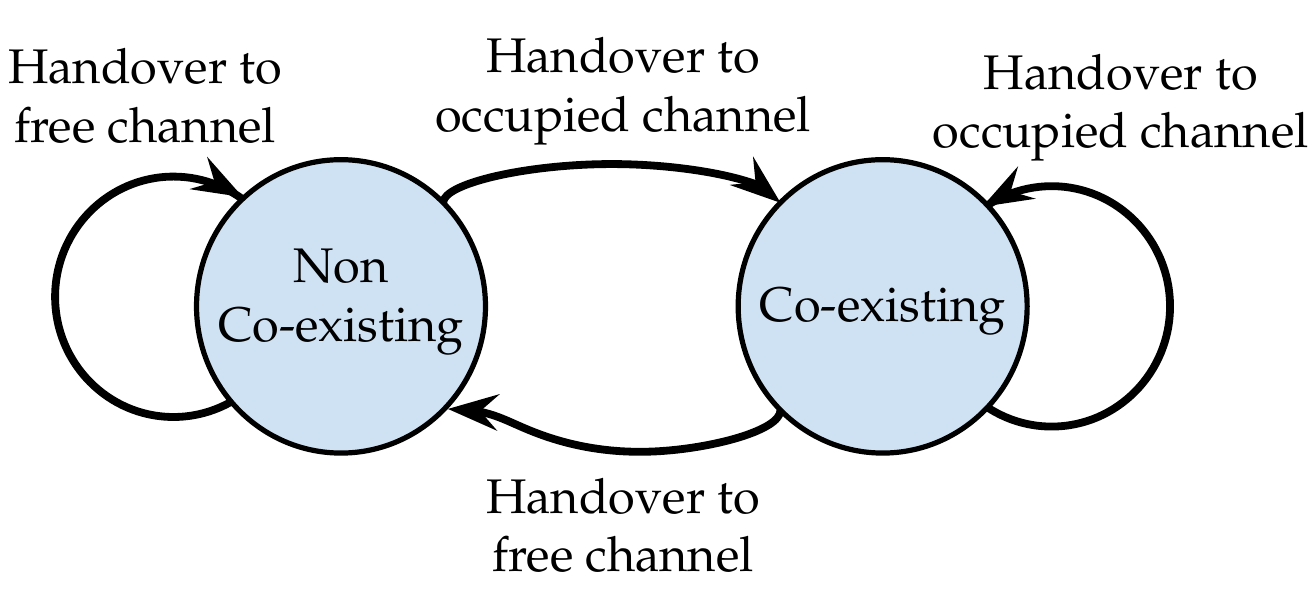}
	%Original: https://docs.google.com/drawings/d/1KB_xZMQWXH-xWfwjSFf2zKc5eLscTIE_jna6T-eQZf8/edit
	\caption{Intra-channel co-existence. Transitions are consequences of handovers.}
	\label{fig:coexistence_automata}
\end{figure}

\Cref{alg:policy} depicts our ranking-based policy function, whose goals are (i) to switch to more favorable frequencies according to the priority defined in \cref{tab:policy}, and (ii) activate the LTE or WiFi co-existing mode if switching to an already occupied frequency. The activation/de-activation of the co-existing mode is depicted in \cref{fig:coexistence_automata}.

Note that we purposely avoid to switch to a frequency whose incumbent technology is unknown to avoid unpredictable communication results.
At line 2 of \cref{alg:policy}, we determine the current classification for the frequency we use for the communication.
Since the \ac{DNN} classifies interference with the \textit{unknown} class, as soon as our system detects unknown or WiFi communications on the currently used frequency (while in non co-existing mode), it reacts by switching channel, possibly switching also to co-existing mode.
Conversely, if \FW is in co-existence, and it detects a clear channel, it swiftly performs a handover to occupy it.
Even if the framework includes functions for tweaking \ac{BS} gain and bandwidth, we choose to use standard values for our experiments, as their use would unnecessarily complicate the policy logic for the purpose of this work.

\begin{algorithm}[!h]
\footnotesize
\caption{Ranking-based policy used in the experiments.}
	\label{alg:policy}
\begin{algorithmic}[1]

\Procedure{rank\_policy}{$channel\_classes,curr\_freq$}
	\State $curr\_class \leftarrow get\_class(channel\_classes,curr\_freq)$
	\If{$curr\_class = (\text{WiFi}|\text{Unknown})$}
	\State $freq,class \leftarrow \max_{\cref{tab:policy}}(channel\_classes)$
	\If{$coexisting$} \Comment{currently in co-existence}
		\If{$class = \text{CLEAR}$}
		\State return($freq,FALSE,std\_gain,std\_bw$)
		\EndIf
	\Else \Comment{new interference detected}
		\If{$best\_class = \text{CLEAR}$}
		\State return($freq,FALSE,std\_gain,std\_bw$)
		\ElsIf {$best\_class \neq \text{UNKNOWN}$}
		\If {$best\_class = \text{LTE}$}
		\State return($freq,LTE,std\_gain,std\_bw$)
		\Else
		\State return($freq,WiFi,std\_gain,std\_bw$)
		\EndIf
		\EndIf
        \EndIf
        \EndIf
    \State return($curr\_freq,coexisting,std\_gain,std\_bw$)
\EndProcedure

\end{algorithmic}
\end{algorithm}

To classify unknown classes, we employ a classification mechanism for \textit{abstain class} called \textit{\ac{ES}}.~\ac{ES} is the simplest way to compute an uncertainty score for a prediction, by evaluating the entropy of the predicted probability.
Our \ac{DNN} outputs three numbers, which represent the probabilities of the input data to belong to, respectively, clear, LTE or WiFi channels. Let these probabilities be $p_0,p_1,p_2$ respectively, then the entropy is defined as:
\begin{equation}
	H=-\sum_{i=0}^2 p_i\log p_i
\end{equation}
$H$ represents the uncertainty score, lower values mean our \ac{DNN} is more confident of the classification.
Validation of the model allows the selection of a hyper-parameter $\alpha$, and the classification is ultimately defined by:
\begin{equation}
	\text{class} = 
	\begin{cases}
		\arg\max_{i=0,1,2} p_i,\ \text{if}\ H< \alpha\\
		3,\ \text{otherwise}.
	\end{cases}
\end{equation}
Where $0,1,2,3$ represents respectively the clear, LTE, WiFi and unknown classes.  We implemented our prototype by leveraging srsRAN (https://www.srsran.com/), an extension of srsLTE \cite{gomez2016srslte}. We consider four frequencies (5.18, 5.20, 5.22, 5.24 GHz), equally spaced by 20 MHz, which coincide with the channels in the LTE band 46 and Wi-Fi channels. For this reason, we select 20 MHz as the bandwidth of the sensing radio.  We extended the interface of srsRAN to support two additional commands (i) change the frequency of a specific cell; (ii) trigger the handover of the \ac{UE}s from one specific cell to another.

\subsection{Colosseum Modifications and Data Collection}
\label{sec:dataset}

Training a \ac{DNN} requires labelled ground-truth data that is realistic and as less affected by interference as possible. For this reason, we leveraged the Colosseum testbed to meet both requirements. In Colosseum, \ac{COTS} software can be deployed and run remotely; at the same time, the radio frequency channel is emulated, and real-world wireless communications can be elaborated while being protected from interference. Thanks to the \ac{MCHEM}~\cite{chaudhari_scalable_2018}, Colosseum is a large-scale wireless network emulator, originally designed and deployed to support DARPA's spectrum collaboration challenge in 2019. Colosseum servers and USRP \ac{SDR}s allow researchers to experiment with wireless software and protocol stacks; in particular, Colosseum has already been employed for mobile networking with srsRAN~\cite{bonati2021scope}.

\begin{figure}[!h]
    \centering
    \includegraphics[width=0.9\columnwidth]{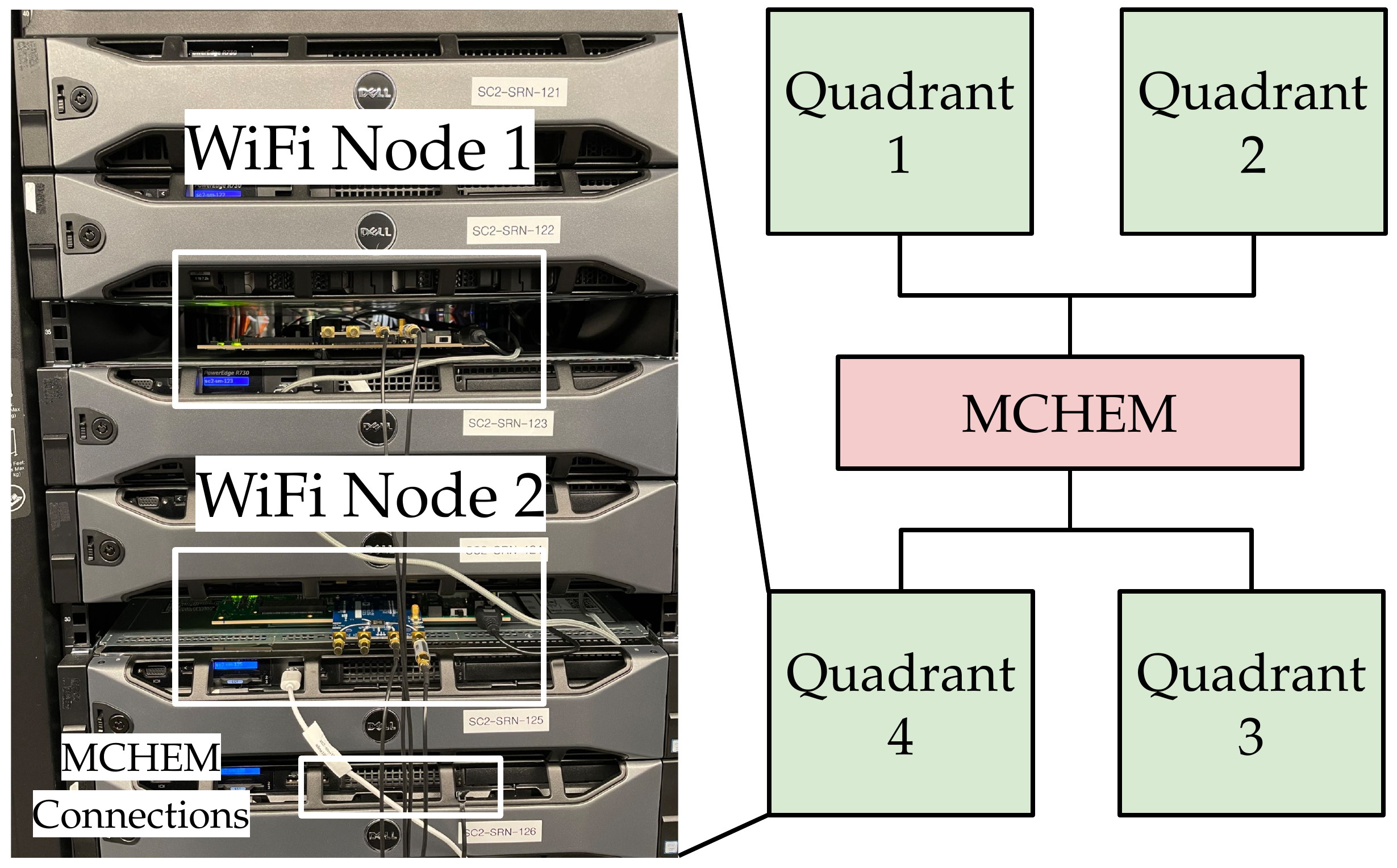}
    \caption{Modifications to Colosseum for OpenWiFi devices.}
    \label{fig:colosseum_wifi}
    % https://docs.google.com/drawings/d/1v_n8z9hrlLRMq-574TDFW8rcT3yfmiunlbFnNxC8Uss/edit?usp=sharing
\end{figure}

On the other hand, the strict timing requirements of 802.11 channel access mechanisms such as CSMA prevent the compliant implementation of the 802.11 stack with \ac{SDR}s~\cite{bloessl_ieee_2013}. In particular, 802.11 requires the reception of an acknowledgement packet for each frame sent, within 10 $\mu$s from the frame successful transmission. To this end, we leveraged a Xilinx ZC706 \cite{ZC706}, a \ac{SoC} board fully supported by the OpenWifi project \cite{jiao2020openwifi}. We worked with the Colosseum management team to deploy a hardware extension of Colosseum for working with WiFi nodes, depicted in Figure \ref{fig:colosseum_wifi}. This extension opens up new experiment opportunities for the whole research community working with 802.11 and spectrum sharing.  The combination of the ZC706 and OpenWifi allows  Colosseum to support fully compliant WiFi devices and communications.

As far as data collection is concerned, we collected three groups of spectrum data, namely background noise (clear), LTE data traffic, and WiFi data traffic. The data captures the general characteristic of LTE and WiFi transmissions, abstracting from the actual transmitted information, and throughput. We used srsRAN for cellular communications and OpenWifi for 802.11. We collected four classes of data:

\begin{enumerate}
	\item Network with Idle traffic;
	\item Continuous high-throughput traffic (iperf3, 1Mbps);
	\item bursty high-throughput traffic (ping flooding, 1KB size);
	\item bursty low-throughput traffic (ping, packets of 300 bytes).
\end{enumerate}

The first class of data is meant to allow the \ac{DNN} to learn of possibly idle base stations or access point that are not transmitting, but that could be potentially impacted by \FW activity. The second and third class of data are meant to represent generic transmissions of random data. The fourth class is similar to the third, and it is used only for experiments as an evidence that \FW is not over-fitting over a particular class of communication patterns, but it is able to extract the crucial wireless technology characteristics from the I/Q samples. Overall, the collected dataset consists of 172.8 GB of data, representing 43.2 billions of I/Q samples, and 18 minutes of communications. Note that we are not collecting samples for the ``unknown'' class.
During data collection, we configured Colosseum to work at 5.24 GHz, and we used five of its nodes (two for LTE communication, two for WiFi communication, and another for data recording).

\section{Experimental Evaluation}
\label{sec:results}

This section is logically divided in three parts.
First, we present in Section \ref{sec:training} the experiments which have led to the \FW \ac{DNN} design for technology classification. Secondly, we showcase the \FW prototype performance in the controlled Colosseum emulator environment in Section \ref{sec:sensing}. In particular, we detail the main features and behaviours \FW can offer to mobile networking spectrum sharing.
Lastly, we demonstrate the developed prototype of \FW on an over-the-air, real-world environment through the use of the Arena testbed in Section \ref{sec:arena}.

\subsection{DNN Training and Testing Procedures}\label{sec:training}

As far as the training of the \ac{DNN} is concerned, we adaptively stop the \ac{DNN} training iterations whenever no sensible progress is gained for a large number of epochs. At the end, we save the network parameters with the best validation results. As in previous related research~\cite{resnet_modulation}, we find the Adam optimizer to be a stable and effective choice, and we use it through our work. We split our dataset (described in \cref{sec:dataset}) in the following way: 50\% for training, 25\% for validation, and 25\% for testing.\smallskip

\textbf{Test-$\alpha$ dataset and $\alpha$ selection.} The \ac{ES} method employed on the \ac{DNN} for the abstain class requires the parameter selection for $\alpha$ (see \cref{sec:scu}).
We create a new dataset, called test-$\alpha$, consisting of the test set plus a combination of LTE traces, representing LTE interference (the latter accounting for the 25\% of test-$\alpha$). While the test dataset is used to evaluate the accuracy of the resulting \acp{DNN}, the test-$\alpha$ dataset is used to tune the parameter $\alpha$.
Specifically, after we train our model using the training set, checking the accuracy on the validation set, we compute the value of $\alpha$ which grants the higher accuracy score on the test-$\alpha$ dataset, and we use it consistently with our model when evaluating the test set.\smallskip

\textbf{Model Selection.} Our investigation focuses on residual networks (RNs)  and convolutional neural networks (CNNs). Since our prototype senses a bandwidth of 20 MHz, it receives a stream of 20M I/Q samples per second from the sensing radio. Therefore, when the \ac{DNN} input size is 2,000, it represents one tenth of millisecond of communication, and, when it is 20,000, it represents one millisecond. Table \ref{tab:architectures} shows the two DNN architectures used throughout the paper, which were inspired by the work presented in \cite{o2018over}.

\begin{table}[!h]
	\centering
	\caption{Two DNN network layouts used in this work. \ac{CNN} architecture (A) and \ac{RN} architecture (B).}
	\label{tab:DNN_layout}
	A)
	\begin{tabular}{|l|l|}
	\hline
	Layer & Output dim.\\
	\hline
	Input & 2 x 20000 \\
	Conv (ReLU) & 7 x 20000 \\
	MaxPool & 7 x 10000 \\
	Conv (ReLU) & 7 x 10000 \\
	MaxPool & 7 x 2000 \\
	Conv (ReLU) & 7 x 2000 \\
	MaxPool & 7 x 1000 \\
	Conv (ReLU) & 7 x 1000 \\
	MaxPool & 7 x 200 \\
	Conv (ReLU) & 7 x 200 \\
	MaxPool & 7 x 100 \\
	Conv (ReLU) & 7 x 100 \\
	MaxPool & 7 x 20 \\
	Conv (ReLU) & 7 x 20 \\
	MaxPool & 7 x 10\\
	FC/Tanh & 18 \\
	FC/Tanh & 16 \\
	FC/Softmax & 3 \\
	\hline
\end{tabular}
~~B)
	\begin{tabular}{|l|l|}
	\hline
	Layer & Output dim.\\
	\hline
	Input & 2 x 20000 \\
	ResidualStack & 4 x 10000 \\
	ResidualStack & 4 x 2000 \\
	ResidualStack & 4 x 1000 \\
	ResidualStack & 4 x 200 \\
	ResidualStack & 4 x 100 \\
	ResidualStack & 4 x 20 \\
	ResidualStack & 4 x 10 \\
	FC/Tanh & 16 \\
	FC/Tanh & 16 \\
	FC/Softmax & 3 \\
	\hline
\end{tabular}
\label{tab:architectures}
\end{table}

We first evaluate how CNN and RN approaches perform against each other. \Cref{fig:input_accuracy} shows the trade-off between input size and achieved accuracy by \ac{RN} and \ac{CNN}.
In these experiments, we vary the number of hidden layers of the models according to the input size to grant always the finest degree of analysis (i.e., we do not increase the kernel size of the convolutional layers). During these experiments, \acp{RN} perform consistently better than \acp{CNN} and can reach an accuracy of 99\% on our validation set. As expected, higher accuracy corresponds to a larger input size. We notice that 20,000 samples is representative enough to reach almost perfection on the validation set, without excessively impacting the processing time (sensing time is 1 millisecond, and processing time in the same order of magnitude).
For the sake of clarity,  the results shown here relate to \acp{RN} and \acp{CNN} with comparable number of parameters (about 3,000). The two network architectures achieving the highest accuracy are shown in \cref{tab:DNN_layout}.

After selecting the network architecture and training the models, we tune the $\alpha$ parameter. \Cref{fig:entropy_tresh} presents an analysis for parameter selection on the previously most successful models. The \ac{RN} model achieves the best results with $\alpha=0.7$, while the \ac{CNN} model requires $\alpha=0.9$. This difference means that \ac{RN} is more confident on its prediction, hence, it requires less account for uncertainty. Overall, the \ac{RN} model obtains better performance and, for this reason, we select \ac{RN} as our architecture of choice for \FW. \Cref{fig:resnet_layer} shows the performance variation when keeping input size to 20,000 while varying the number of layers, meaning opting for larger convolutional kernels. Results show there is little appreciable variation, except for the number of parameters, which increases with larger kernels. Our best \ac{RN} model (architecture shown in \cref{tab:DNN_layout}-B, $\alpha=0.7$) scores an accuracy of 96.4\% on the test set. \Cref{tab:DNN_conf_matrix} shows the resulting confusion matrix for our model, and \cref{tab:DNN_measures} confirms that it is not biased toward any class.

\stdimg{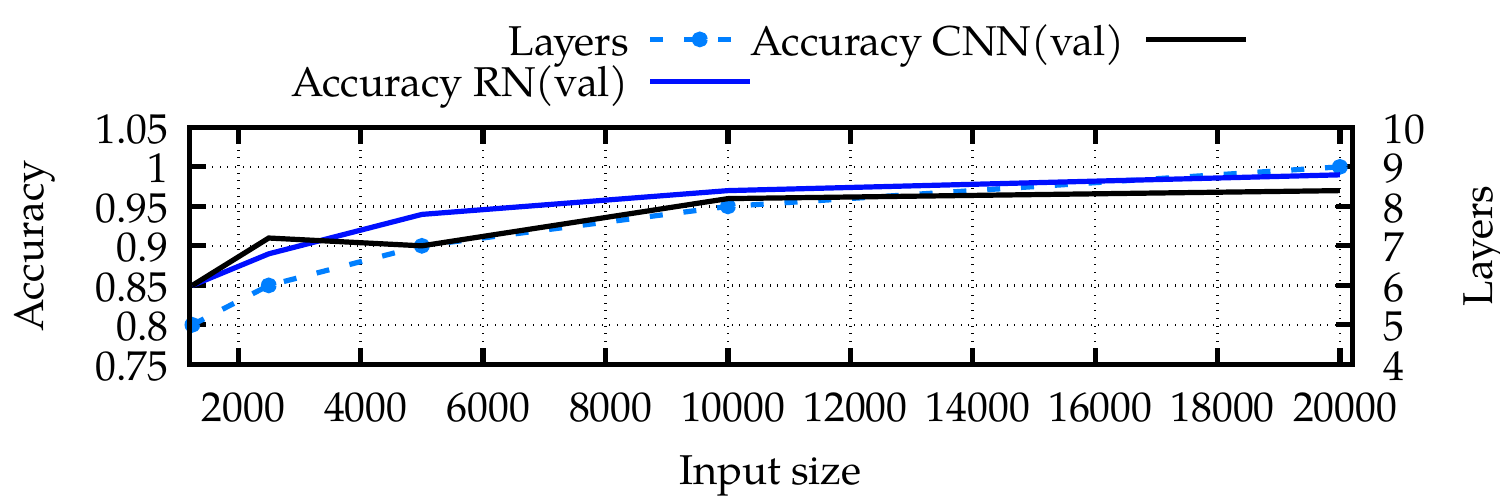}{fig:input_accuracy}{Accuracy on the validation set for \ac{RN} and \ac{CNN} networks varying the input size.}
\stdimg{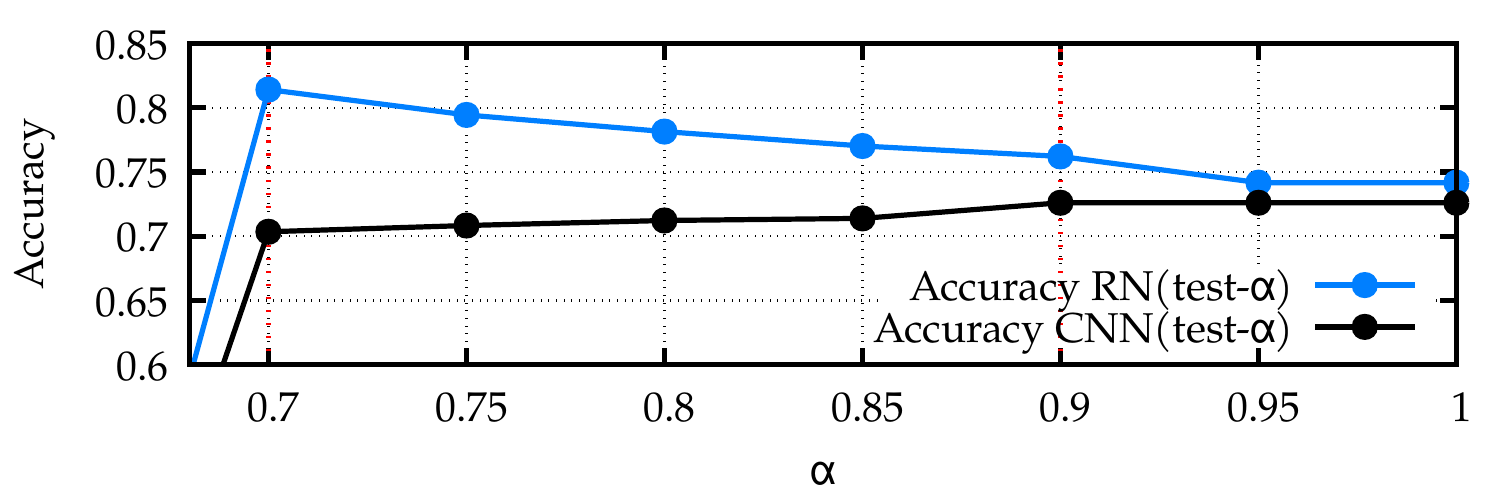}{fig:entropy_tresh}{Accuracy on the test-$\alpha$ set for \ac{RN} and \ac{CNN} models varying$\alpha$. Red lines intersect the maxima.}
\stdimg{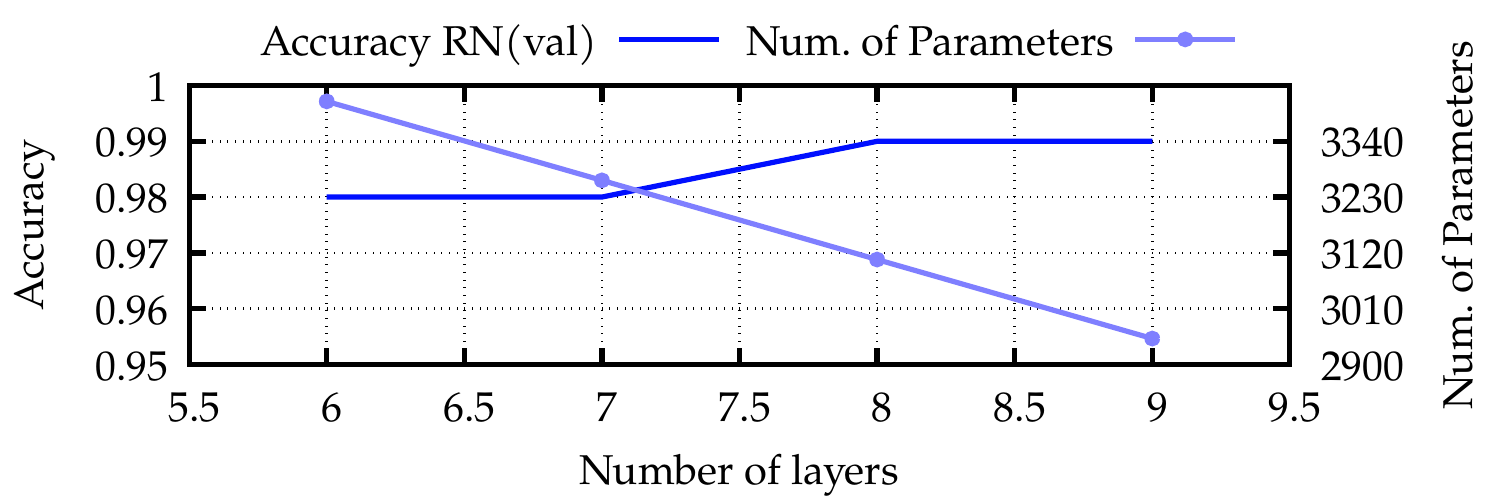}{fig:resnet_layer}{Accuracy on the validation set for \ac{RN} models varying the number of hidden layers.}

\begin{table}
	\centering
	\caption{Confusion matrix of our \ac{DNN} on the test set.}
	\label{tab:DNN_conf_matrix}
    \scriptsize
	\begin{tabular}{|l|l|l|l|l|}
		\hline
		&Clear & LTE & WiFi & Unknown\\
		\hline
		%Clear & 8.7202e+04& 2.0000e+00& 1.1000e+02& 2.6860e+03\\
		%LTE & 3.0000e+00& 8.8908e+04& 1.0400e+02& 9.8500e+02\\
		%WiFi & 5.7700e+02& 1.5200e+02& 8.4066e+04& 5.2050e+03\\
	    Clear & 87,202 (32.3\%) & 2 (0\%) & 110 (0\%) & 2,686 (1\%)\\
		LTE & 3 (0\%) & 88,908 (32.9\%) & 104 (0\%) & 985 (0.4\%)\\
		WiFi & 577 (0.2\%) & 152 (0.1\%) & 84,066 (31.1\%) & 5,205 (1.9\%)\\
		\hline
\end{tabular}
\end{table}
\begin{table}
	\centering
	\caption{Recall, Precision and F1 measures for our \ac{DNN} on the test set.}
	\label{tab:DNN_measures}
	\begin{tabular}{|l|l|l|l|}
		\hline
		Technology & Recall & Precision & F1 \\
		\hline
		\hline
		Clear & 0.9689 & 0.9934 & 0.981\\
		\hline
		LTE & 0.9879& 0.9983& 0.9931 \\
        \hline
		WiFi & 0.9341& 0.9975& 0.9648 \\
        %\hline
	%	Unknown & nan& 0.0& nan \\
        \hline
\end{tabular}
\end{table}

\subsection{Sensing and Reacting}\label{sec:sensing}

In this section, we demonstrate the spectrum optimization capabilities of \FW. Specifically, we test \FW in a controlled environment, so to be able to (i) emulate corner spectrum conditions, and (ii) obtain repeatable results. We leverage Colosseum since it enables fine-grained control over the wireless environment. In this section, we demonstrate that:

\begin{enumerate}[(a)]
	\item \FW can detect interference with its communication,
	\item it can perform a handover of the existing mobile communication to a difference frequency,
	\item it changes its mode to \textit{co-existence} if switching to an already occupied frequency, and
	\item its choices of frequency and whether to enable co-existence are close to the optimum (for a given policy).%its on-going communication is never halted in the process.
\end{enumerate}

The experiments starts with the RU transmitting on the unoccupied 5.18 GHz channel, along with other two transmissions on 5.22 and 5.24 GHz.
LTE and WiFi transmissions were recorded using Colosseum (4th class of the dataset, not used for training/validating/testing of the model) and are used in place of real nodes to make the experiment finely-controlled and reproducible. During the duration of the experiment, we keep an active communication between the \FW \ac{BS} and a srsRAN \ac{UE} using a ping session. We log such continuous communication to check whether data is lost due to interference and handovers. We conduct extensive experimentation using Colosseum, varying the number of nodes and transmissions to stress \FW.
Here, we present results from two sessions that highlight all \FW aspects, demonstrating claims (a)-(d).

\begin{figure}[!h]
    \centering
    \includegraphics[width=\columnwidth]{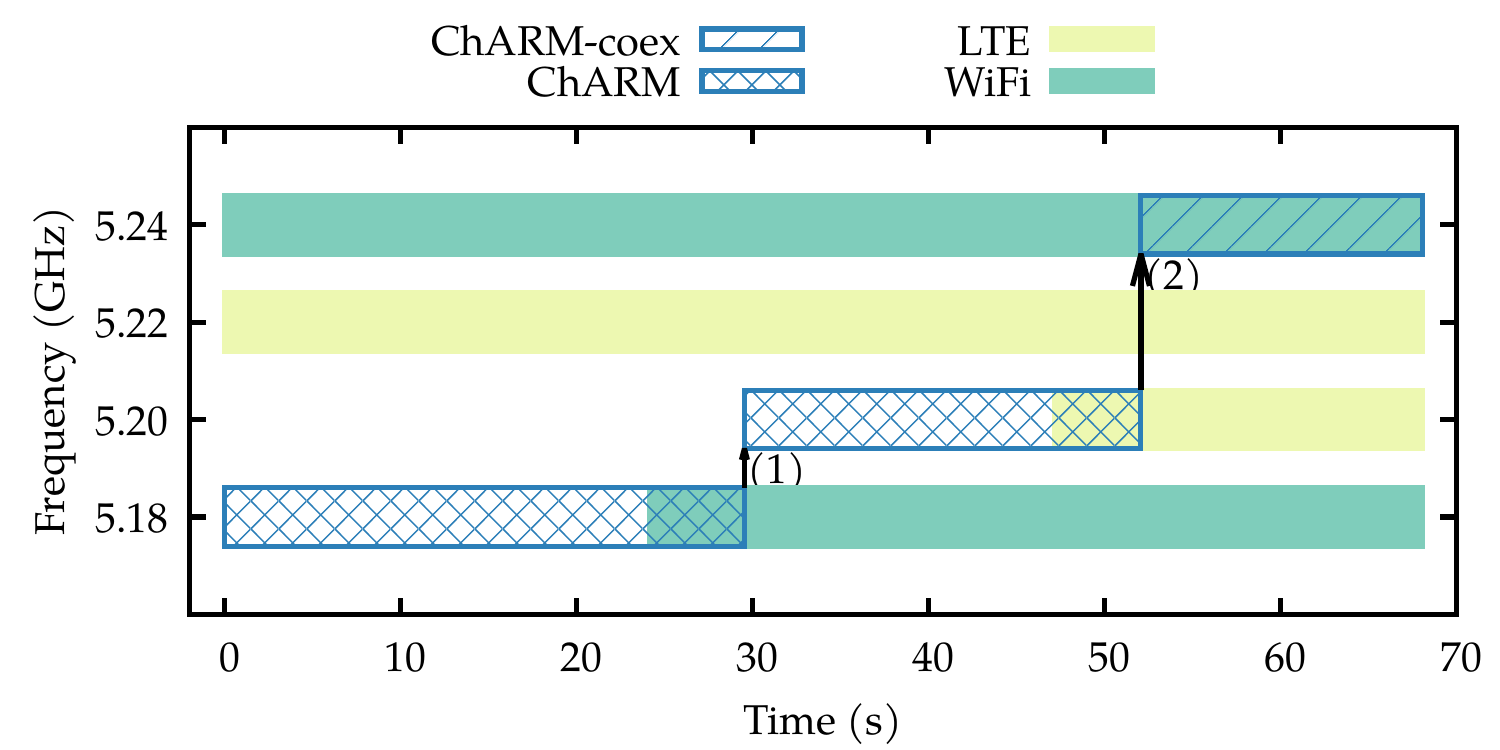}
    \caption{Frequency occupancy during an experiment on Colosseum. Reported transmission classes are the ground-truth. Arrows indicate the triggering of handovers. The policy used is based on \cref{tab:policy}-A}
    \label{fig:charm}
\end{figure}

\Cref{fig:charm} shows the channel transmissions during one of the experiments. Around the 24th second, an interfering WiFi transmission starts at frequency 5.18 GHz, the same that \FW is using. It takes a few seconds -- i.e., the overlapping boxed area in \Cref{fig:charm} -- for \FW to detect interference and trigger a handover, indicated by arrow number 1. The detection of interference demonstrates claim (a), while the 3GPP standardized inter-cell handover guarantees  continuity of the data communication between \FW and the attached \ac{UE}, demonstrating (b). \FW handovers to frequency 5.2 GHz, which according to the policy defined by \cref{tab:policy}-a, is the best choice, and does not require a co-existence mechanism. Around the 52nd second, an interfering LTE transmission starts on the same frequency as \FW.
After a few seconds -- shown in \cref{fig:charm} with the second overlapping area --  \FW correctly detects interference with another LTE network and triggers a handover, indicated by arrow number 2. Claims (a) and (b) are further demonstrated, as \FW detects the interference (a) and seamlessly perform a 3GPP handover (b). In this case, since there are no unoccupied frequencies among those under evaluation, \FW follows the policy in \cref{tab:policy}-a and switches to 5.24 GHz. Since \FW  detects an existing WiFi communication there, it activates the co-existence mode, hence demonstrating claim (c). It is worth noting that the choices performed by \FW are optimal with respect to the policy. Besides the intervals for which the delay in channel sensing prevents the correct classification of channel occupancy, \FW detects the correct underlying traffic, it switches to the expected frequency given the policy, and it activates the appropriate co-existence mode when needed, demonstrating claim (d).\vspace{-0.2cm}

\begin{figure}[!h]
    \centering
    \includegraphics[width=\columnwidth]{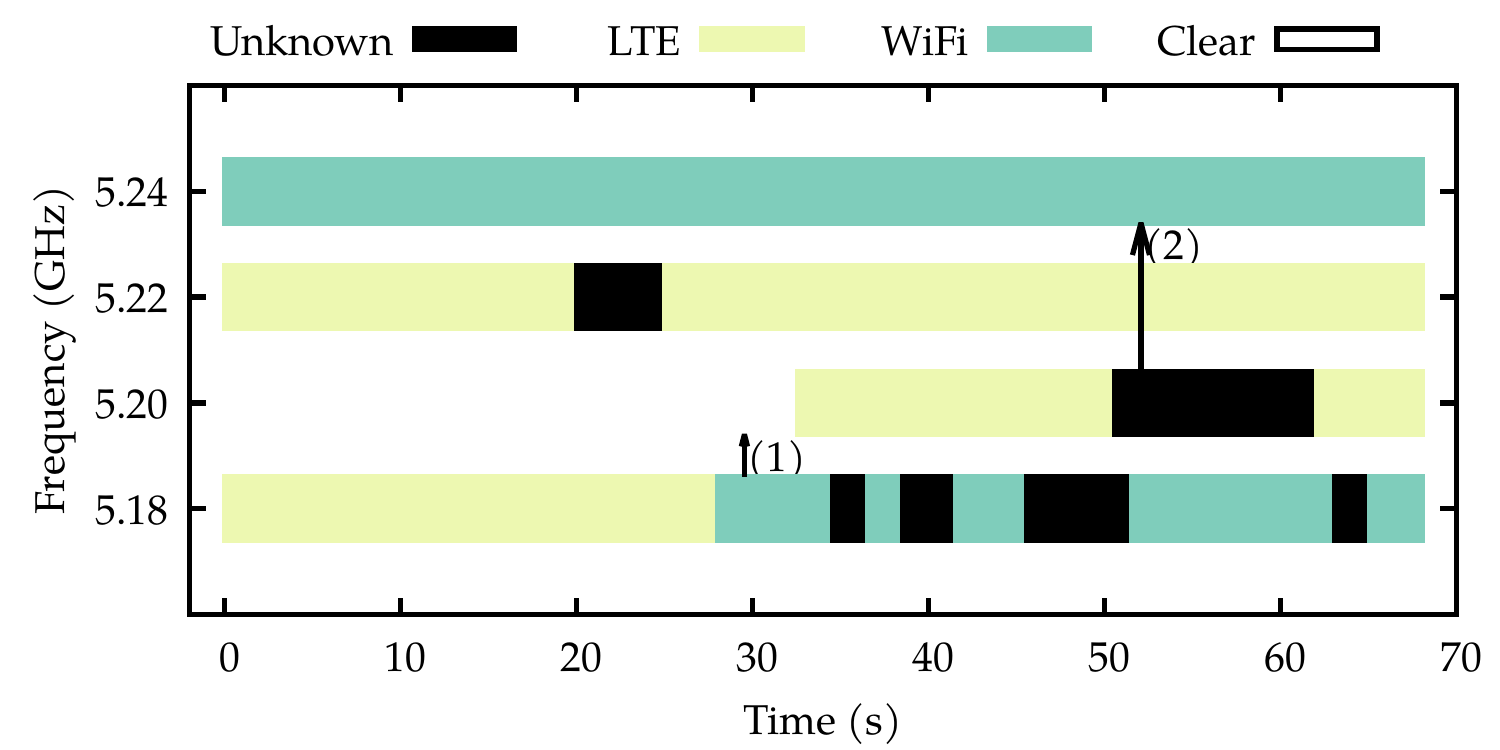}
    \caption{\FW \ac{DNN} frequency classification through the experiment. Arrows indicate the triggering of handovers.}
    \label{fig:charm_class}
\end{figure}

While \cref{fig:charm} presents the objective development of the experiment, we show in \cref{fig:charm_class} the classification outcomes of the DNN throughout the experiment. Specifically, we notice the interference detection, determining the triggering of the handovers (1,2), and some misclassification of LTE and WiFi communications as unknown technologies.

\begin{table}[!h]
	\centering
	\caption{Confusion matrix for \FW classification during an experiment.}
	\input{script/charm_conf_mat}
	\label{tab:exp_conf_mat}
	\end{table}

\Cref{tab:exp_conf_mat} presents the performance of the \ac{DNN} during the experiment in terms of classification confusion matrix.
As expected, the main source of misclassification stems from uncertainty in the interference and the abstain classes. \Cref{fig:charm_alt} presents another run showing the flexibility of \FW, where \cref{tab:policy}-b is enacted. In the spirit of O-RAN networks, we demonstrate that a simple change in the policy determines the preference of \FW for co-existing with LTE technologies in place of WiFi, without any significant structural change.

\stdimg{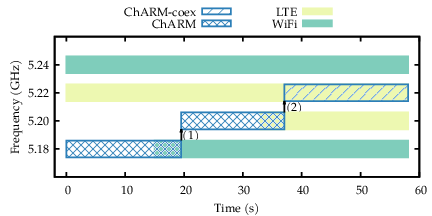}{fig:charm_alt}{Frequency occupation during an experiment on Colosseum. Reported transmission classes are the ground-truth. Arrows indicate the triggering of handovers. The policy used is based on \cref{tab:policy}-B.}

\subsection{Over-the-air Experimental Evaluation on Arena}\label{sec:arena}

\begin{figure}[!h]
    \centering
    \includegraphics[width=\columnwidth]{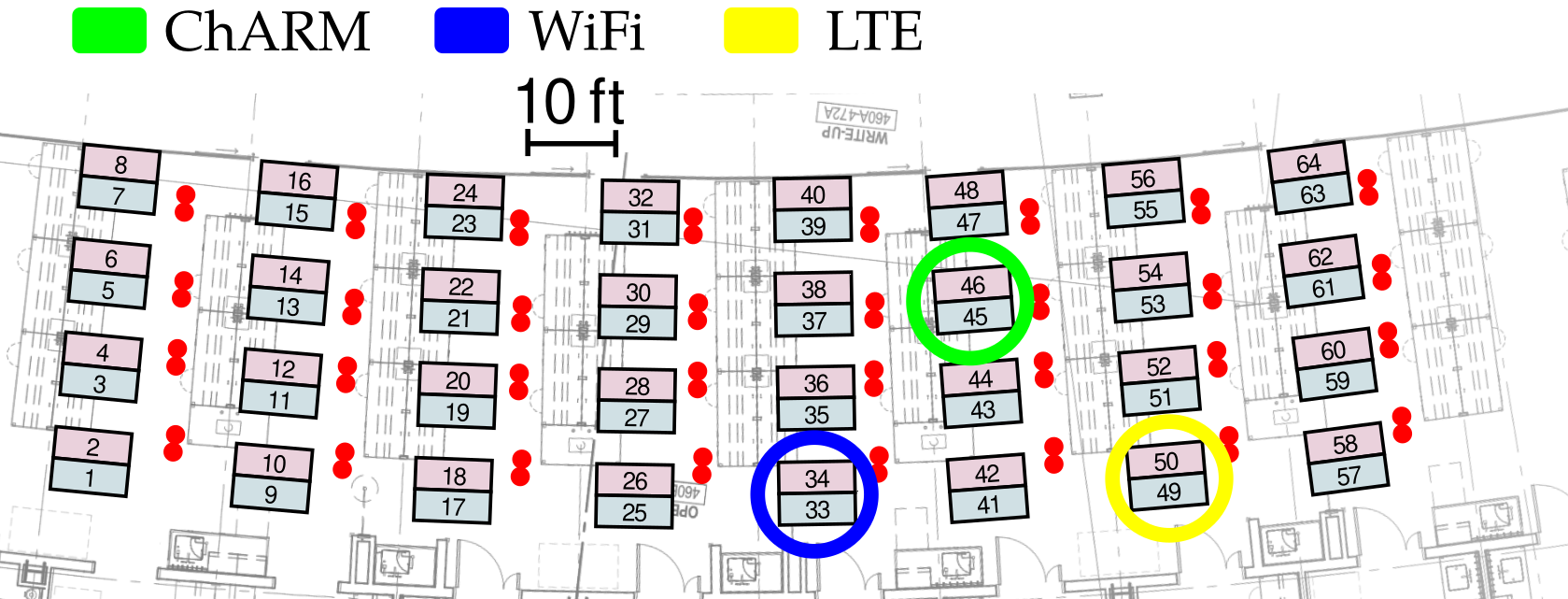}
    \caption{Location of the wireless nodes in the Arena testbed.}
    \label{fig:arena}
\end{figure}

We leveraged the Arena testbed \cite{bertizzolo_arena_2020} to perform over-the-air testing of \FW. Arena is a remotely accessible and open test-bed made up of 64 \acp{SDR} and 8 servers designed for experimenting with 5G-and-beyond spectrum research. It allows the deployment and testing of communication platforms in a real office environment during working hours, subject to all sort of \ac{ISM} interference. The antennas are installed in a large shared office, displaced along a grid, as represented in \cref{fig:arena}. The positioning in the grid of the nodes used in the experiments is also shown in the figure.

While Colosseum allows experimenting in a fully controlled environment, Arena allows the testing of our system in a more realistic and challenging environment. \Cref{fig:arena} presents the deployment of \FW in Arena, including the location of the LTE and WiFi transmitting nodes. We choose 5.24 GHz as the center frequency. To generate the LTE and WiFi traffic, we use the traffic traces from our dataset. While the experiments on Colosseum show the behavior of \FW and its ability to optimize the spectrum use, the experiments on Arena validate that the \acp{DNN} are still effective in a real-world environment. We emphasize that no data has been collected on Arena and used for the training of our model.

To compensate the path loss in Arena with respect to Colosseum, the transmissions are amplified by 10 dB. For the targeted LTE and WiFi transmission configurations (in terms of bandwidth parameters, LTE physical channel allocation, and WiFi modulation scheme) and transmission power, results shown in \cref{fig:charm_arena} confirm that \FW can reliably detect the technologies obtaining an accuracy of 85\%.
Variations in the transmission configuration or abrupt changes in the transmission power can however reduce our classifier performance, as shown in \cref{fig:power_measures}. 
Thus, it is important to tune the \ac{DNN} training with respect to the target environment characteristics, and future work will investigate how online training can be used to automatically tailor the performance of \FW.

\stdimg{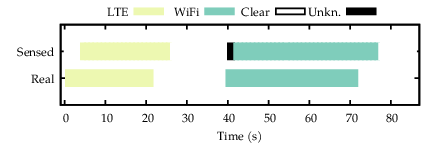}{fig:charm_arena}{Sensing experiment on the Arena test-bed at 5.24 GHz, with 20 MHz of bandwidth. We play out LTE and WiFi communication recordings (indicated as \textit{Real} transmissions) and verify \FW classification (\textit{sensed}).}
\stdimg{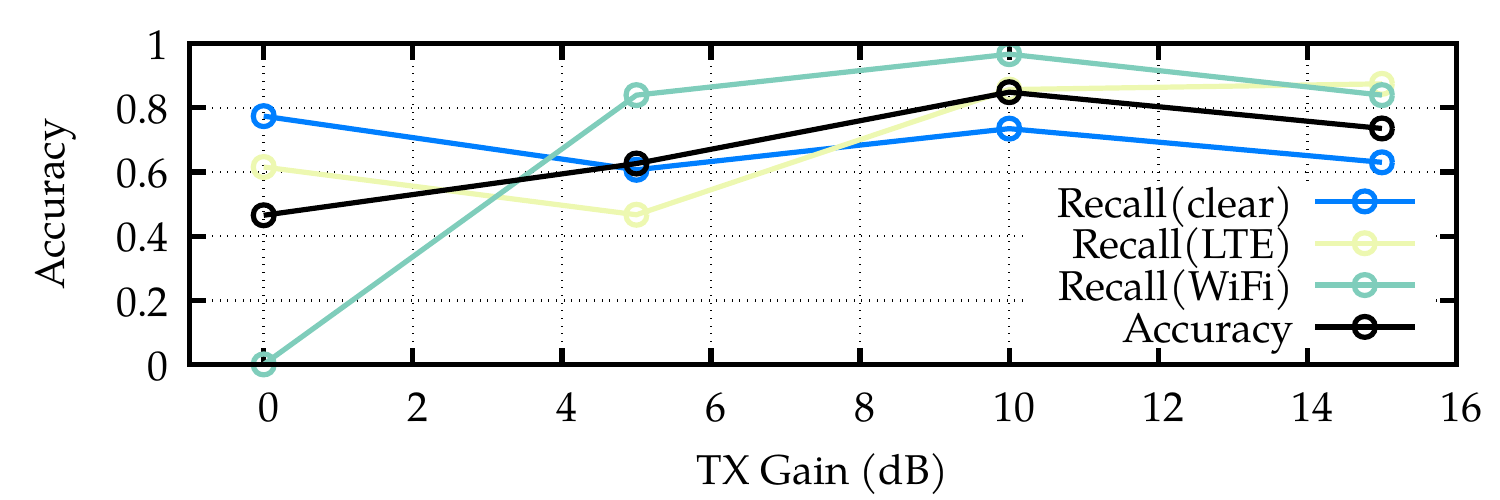}{fig:power_measures}{Accuracy and recall of \FW \ac{DNN} obtained in the wireless test-bed for a single channel (5.24 GHz). The TX Gain is wrt the original transmission on Colosseum.}

\section{Conclusions}
\label{sec:conclusions}
In this paper, we have proposed Channel-Aware Reactive Mechanism (\FW), a data-driven O-RAN-compliant framework that allows (i) sensing of the spectrum to understand the current context and (ii) reacting in real time by switching the distributed unit (DU) and RU operational parameters according to a specified spectrum access policy. It is designed to operate within the O-RAN specifications, and can be used in conjunction with other spectrum sharing mechanisms. We demonstrate the performance of \FW in the context of spectrum sharing among LTE and Wi-Fi in unlicensed bands, where a controller operating over a RAN Intelligent Controller (RIC) senses the spectrum and switches cell frequency to avoid Wi-Fi. We develop a full-fledged standard-compliant prototype of \FW using srsRAN, and leverage the Colosseum channel emulator to collect a large-scale waveform dataset to train our neural networks with.
Experimental results show that our neural networks achieve accuracy of up to 96\% on Colosseum and 85\% on Arena, demonstrating the capacity of \FW to fully exploit the considered spectrum channels. 

The authors have provided public access to their code at https://github.com/lucabaldesi/charm\_code  and to their dataset at http://hdl.handle.net/2047/D20423481

\bibliographystyle{IEEEtran}
\bibliography{biblio}
\end{document}

%% file: script/charm_conf_mat.tex
\begin{tabular}{|l|l|l|l|l|}
	\hline
	&Clear & LTE & WiFi & Unknown\\
	\hline
	Clear & 30 (11.2\%) & 0 (0.0\%) & 0 (0.0\%) & 0 (0.0\%) \\
	LTE & 2 (0.7\%) & 106 (39.6\%) & 0 (0.0\%) & 14 (5.2\%) \\
	WiFi & 0 (0.0\%) & 0 (0.0\%) & 76 (28.4\%) & 13 (4.9\%) \\
	Intrf & 0 (0.0\%) & 8 (3.0\%) & 17 (6.3\%) & 2 (0.7\%) \\
	\hline
\end{tabular}